\documentclass[a4paper,11pt]{article}
\usepackage{pos}

\title{Effective pointing of the ASTRI-Horn telescope using the Cherenkov camera with the Variance method}
\ShortTitle{Effective pointing of ASTRI-Horn with the Variance method}

\manuallySeparateAuthors
\author*[a,b]{Simone~Iovenitti,}
\author[b]{ Giorgia~Sironi,}
\author[c]{ Alberto~Segreto,}
\author[c]{ Osvaldo~Catalano}
\author[c]{ and~Teresa~Mineo,}
\author[d]{ for the ASTRI Project}

\affiliation[a]{Università degli Studi di Milano, dip.to di Fisica,\\
Via Giovanni Celoria 16, 20133 Milano, Italy}

\affiliation[b]{INAF--Osservatorio Astronomico di Brera,\\
Via E. Bianchi 46, 23807 Merate, Italy}

\affiliation[c]{INAF--IASF Palermo,\\
Via U. La Malfa 153, 90146 Palermo, Italy}

\affiliation[d]{\protect\url{astri.inaf.it}}

\emailAdd{simone.iovenitti@inaf.it}

\abstract{Cherenkov telescope cameras are not suitable to perform astrometrical pointing calibration since they are not designed to produce images of the sky, but rather to detect nanosecond atmospheric flashes due to very high-energy cosmic radiation. Indeed, these instruments show only a moderate angular resolution (fractions of degrees) and are almost blind to the steady or slow-varying optical signal of starlight. For this reason, auxiliary optical instruments are typically adopted to calibrate the telescope pointing. However, secondary instruments are possible sources of systematic errors. Furthermore, the Cherenkov camera is the only one framing exactly the portion of the sky under study, and hence its exploitation for pointing calibration purposes would be desirable.
In this contribution, we present a procedure to assess the pointing accuracy of the ASTRI-Horn telescope by means of its innovative Cherenkov camera. This instrument is endowed with a statistical method, the so-called Variance method, implemented in the logic board and able to provide images of the night sky background light as ancillary output.
Taking into account the convolution between the optical point spread function and the pixel distribution, Variance images can be used to evaluate the position of stars with sub-pixel precision. In addition, the rotation of the field of view during observations can be exploited to verify the alignment of the Cherenkov camera with the optical axis of the telescope, with a precision of a few arcminutes, as upper limit. This information is essential to evaluate the effective pointing of the telescope, enhancing the scientific accuracy of the system.
}

\FullConference{37$^{\rm{th}}$ International Cosmic Ray Conference (ICRC 2021)\\
		July 12th -- 23rd, 2021\\
		Online -- Berlin, Germany}

\begin{document}
\maketitle

\section{Introduction}
\begin{figure} %
\centering
    \includegraphics[width=0.5\textwidth]{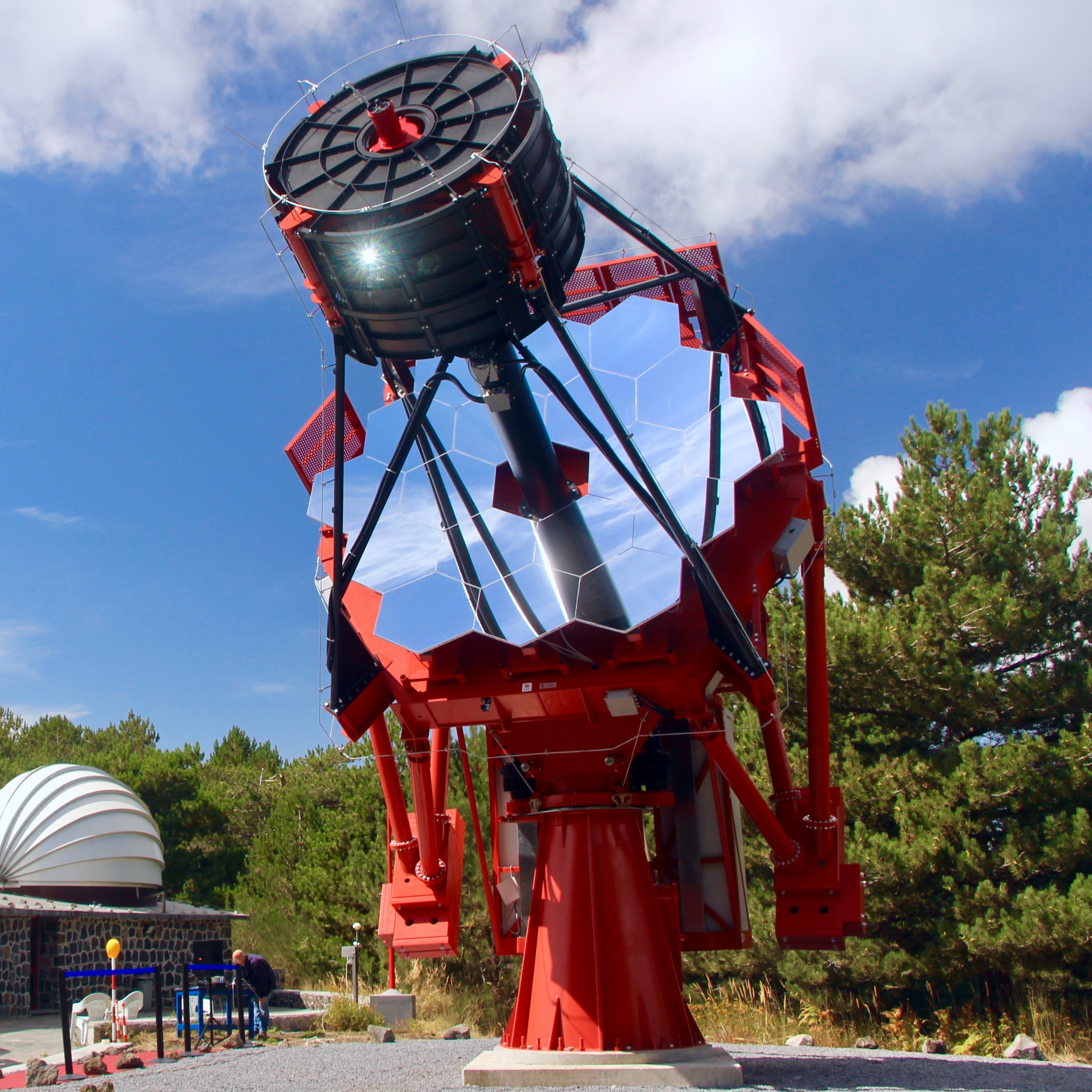}
    \caption{The ASTRI-Horn telescope, located on Mount Etna (Italy). Photo credit: E. Giro.}
\label{fig_1}
\end{figure}

\noindent
ASTRI (acronym of \textit{Astrofisica con Specchi a Tecnologia Replicante Italiana}) is an Italian project aiming at the realization of 9 telescopes (Mini-Array) to be installed in Tenerife (Spain) \cite{Antonelli}. A prototype telescope, ASTRI-Horn, was installed in 2014 on Mount Etna (Italy) and it is currently in operation (figure~\ref{fig_1}) \cite{ASTRI_validation_2019}. It is an innovative example of Imaging Atmospheric Cherenkov Telescope (IACT): an instrument designed to detect the cosmic gamma radiation that reaches our planet, measuring nanosecond light flashes produced in the atmosphere by the Cherenkov effect due to showers of secondary particles.
The ASTRI-Horn telescope is characterized by some innovative features \cite{astri_optical_validation}: a dual mirror optical scheme (modified Schwarschild-Couder design) and a Cherenkov camera made of silicon detectors (SiPMs) instead of more traditional photo-multiplier tubes (PMT). Moreover, the innovative ASTRI Cherenkov camera is equipped with an ancillary output, the so-called “Variance”, which is not dedicated to the detection of Cherenkov events, but rather to the imaging of the night sky. Using the Variance, we can visualize a snapshot of the field of view (FOV) actually framed by the telescope and hence perform astrometric measurements of bright stars (up to the 7th magnitude) directly from the Cherenkov camera, without any additional hardware. This technique would be a great help during the assembly verification phase and the calibration of the instrument pointing accuracy, but it is limited by the coarse angular resolution, as the Cherenkov camera has a pixel size of 7 mm, equivalent to $\sim$11 arcmin. However, using long observing runs in tracking mode, it is possible to enhance the angular precision of the Variance exploiting the FOV rotation effect, as it will be discussed in the next session. In this way we can adopt the Variance sky view to monitor several quantities related to the pointing of the telescope \cite{Segreto_calibration}. In particular, we want to present hereafter a procedure to measure the alignment of the Cherenkov camera with the optical axis of the telescope. This quantity is very difficult to be accessed otherwise, but it is crucial to guarantee an effective pointing of the telescope, optimizing the accuracy of the whole system.

\section{“Variance” sky view and FOV rotation}
\begin{figure} %
\centering
    \includegraphics[width=0.7\textwidth]{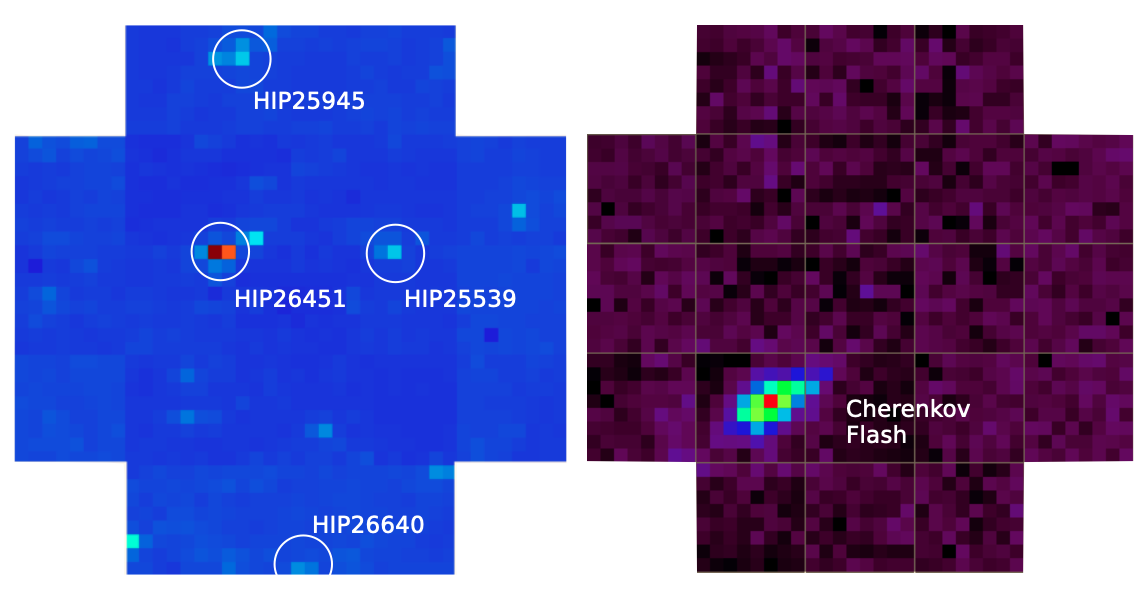}
    \caption{Left: image of the star field available from the Variance output of the camera. The ASTRI-Horn telescope is pointing towards the Crab Nebula (at the center of the FOV). Right: Cherenkov flash recorded on nanoseconds-timescale in the main camera output from the same pointing.}
\label{fig_2}
\end{figure}

\noindent
The Cherenkov camera of the ASTRI telescope is a device dedicated to the detection of the Cherenkov light flashes in the atmosphere. It operates on the time scale of a few tens of nanoseconds, which is the typical duration of Cherenkov events, but it is blind to the continuous light from the night sky background (NSB) and, in particular, to the stars (figure~\ref{fig_2}, right). On the other hand, the ASTRI Cherenkov camera is also equipped with an ancillary output dedicated to the imaging of the NSB, the so-called “Variance” technique: a statistical method implemented in the logic board of the camera, providing a measure of the electromagnetic flux impinging on each pixel in the wavelength range between 300 and 500 nm (the sensitive band of the telescope). Neglecting the dark current, the Variance output of the camera $\sigma^2$ can be expressed by \cite{Segreto_calibration}
\begin{equation}
   \sigma^2 = k \cdot \Phi_{sky}\,\,\,\,  ,
\end{equation}
where $\Phi_{sky}$ is the actual flux from the sky and $k$ is a constant. In the case of the ASTRI-Horn telescope, each Variance image of a given observing run is produced by the back-end electronics  every $\sim$3 s, integrating the pixel signals in absence of shower events for $\sim$1 s ($2^{16}$ samples). In the Variance view of the sky, the star component of the NSB is clearly visible (figure~\ref{fig_2}, left).
\noindent
During long observing runs in tracking mode, the FOV presents the apparent rotation which is typical of every telescope with an alt-azimuth mount. It is due to the evolution of the parallactic angle, i.e. the one between the north celestial pole and the zenith, with vertex in the pointing direction. The FOV rotation can be easily simulated with appropriate software \cite{Iovenitti_star_coverage} and used as a diagnostic tool to monitor the opto-mechanical behaviour of the telescope. In fact, the center of the FOV rotation corresponds to the actual pointing direction: the celestial coordinate individuated by the optical axis of the telescope. In principle, the center of rotation should correspond to the camera geometric center, but mechanical tolerances or gravity flexures could introduce a mis-alignment of the camera mount system, which is difficult to detect with a direct measure. Thanks to the detection of the FOV rotation effect with the Variance, the presence of this possible offset can be investigated and the resulting information can be inserted in the telescope pointing model, enhancing the accuracy of the system. More details on the effectiveness and possible limitations of this method will be provided in a future publication \cite{Iovenitti2021}.

\section{Reconstruction of the star path}
\begin{figure} %
\centering
    \includegraphics[width=0.5\textwidth]{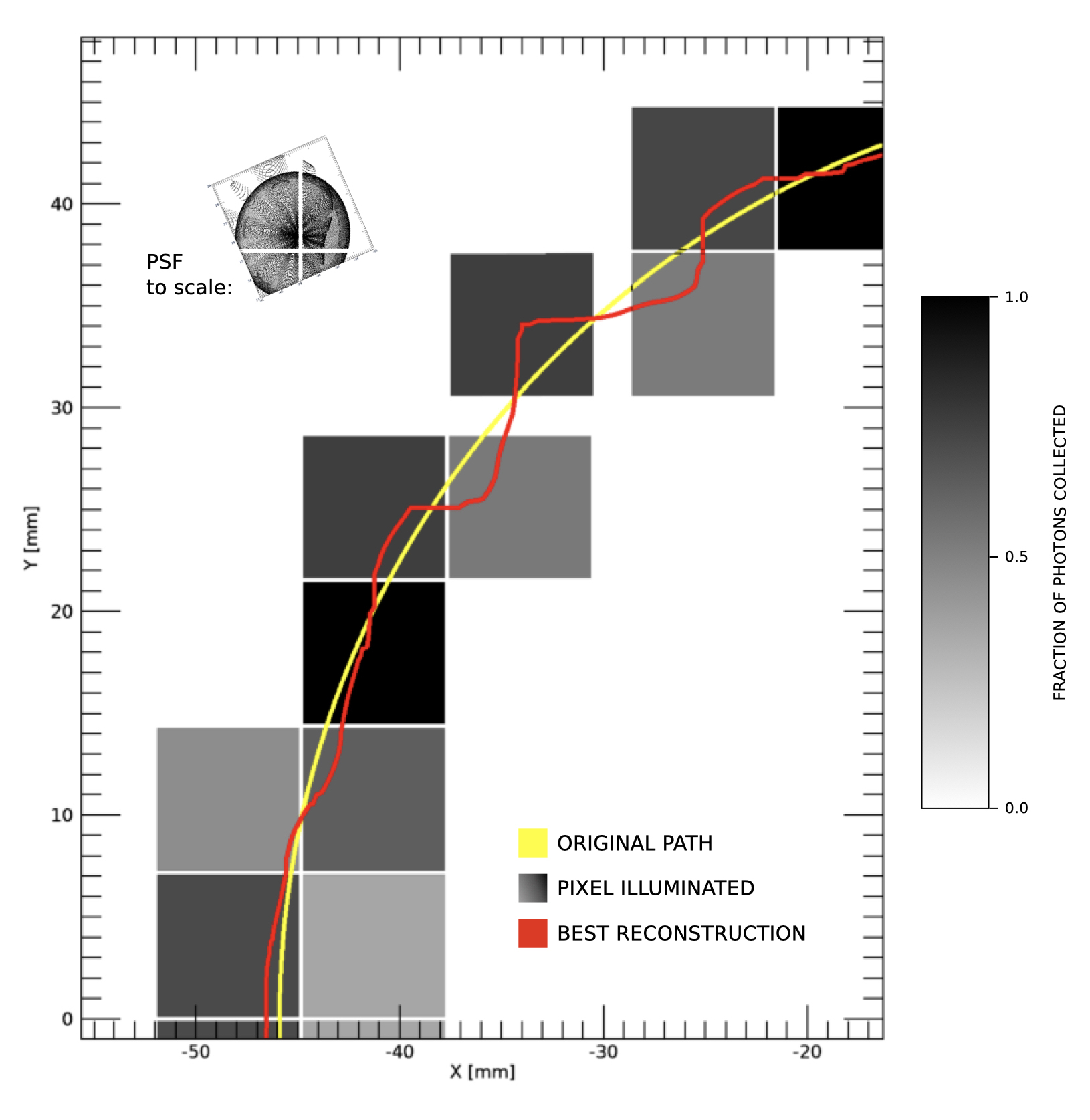}
    \caption{Simulation of a long tracking observing run: the star centroid moves along a circle (yellow) and its light spot is integrated over pixels (grey scale) allowing us to reconstruct the original path (red).}
\label{fig_3}
\end{figure}

\begin{figure} %
\centering
    \includegraphics[width=\textwidth]{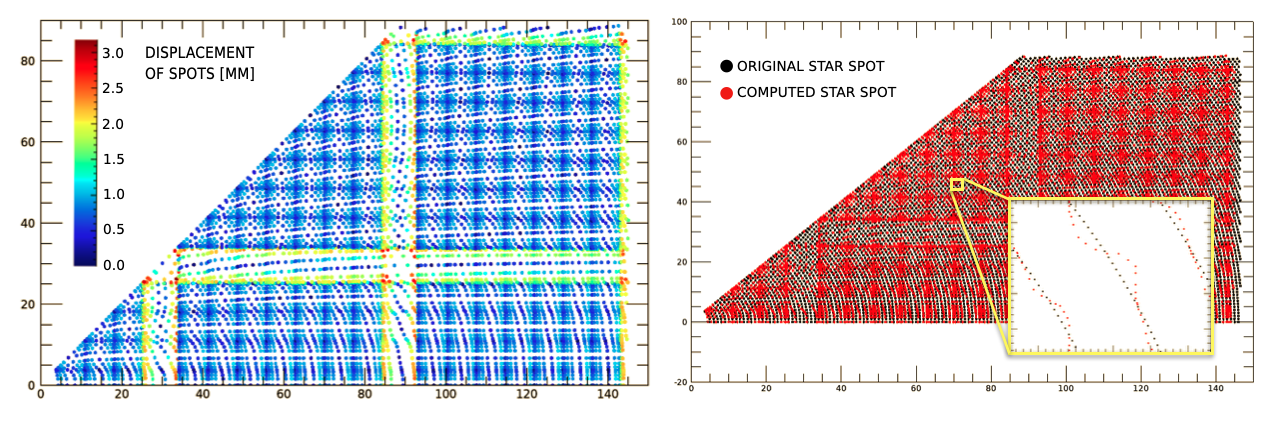}
    \caption{Transformation map (left) and the procedure to compute the displacement of each point (right).}
\label{fig_4}
\end{figure}

\noindent
The main issue in the study of the FOV rotation with Variance data is the reconstruction of the real star path, as the light spot is convolved over the large camera pixelization and hence the information about the star position cannot be directly retrieved. We handled this problem studying the whole process of image creation with specific simulations.\\
We considered an hypothetical long observing run, with a consistent angular rotation of the FOV, and we modeled the light spot of a generic star in the FOV with a custom ray-tracing program (specifically developed by G. Sironi), of about 80000 photons (the size of the point spread function of the telescope is almost equal to the side of the pixel, about 11’). The light was integrated over the pixels, obtaining a “Variance-like” image of the illumination of the camera. Now, considering the intensity recorded in every pixel, we tested different methods to retrieve the original position of the star: the most effective algorithm implements a weighted-average-like process and it provides a reconstruction of the star path with an accuracy of 0.8’ (standard deviation). The complete procedure is shown in figure~\ref{fig_3}.
\noindent
The largest deviation from the original star path is in correspondence with the gap between the photo detection modules (PDMs), the 21 tiles of 64 pixels composing the camera of ASTRI-Horn \cite{camera}. To reduce the distortions introduced by the gaps, we adopted a transformation matrix containing the simulated displacement of spots, sampled over a fine grid of positions in the whole camera. Figure~\ref{fig_4} reports both the procedure to create the matrix (right) and the resulting map of displacements (left). The output of the previous pixel analysis was corrected with this matrix, reducing the dispersion around the original star path of about 40\% (i.e. obtaining $\sim$0.5' RMS).

\section{Multi-ellipse fit procedure}
\begin{figure} %
\centering
    \includegraphics[width=0.8\textwidth]{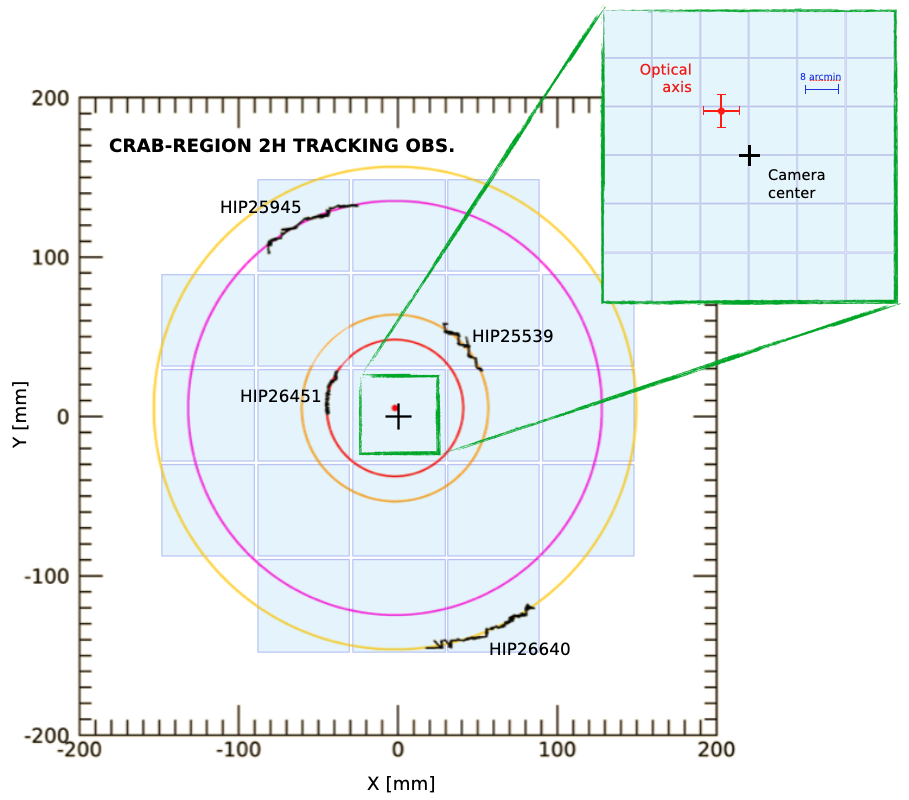}
    \caption{Multi-ellipse fit (coloured circles) of real data (black lines) taken with ASTRI-Horn (RUN1597). The green box shows the results of the analysis on the 3 runs reported in the text.}
\label{fig_5}
\end{figure}

\noindent
Using the simulations reported in the previous section, we validated the procedure and the algorithm to perform the analysis of the Variance images. After that, we looked in the ASTRI-Horn archive for observing runs with real Variance data presenting a consistent FOV rotation (i.e. $>$25°) and 4 stars clearly visible at least. We found out 3 observing runs (ID number 1597, 1605 and 1620, taken respectively on Feb. 26, 27 and 28, 2019), all pointing at the Crab Nebula at the same time and hence containing the same stars in the same position (namely HIP25945, HIP26451, HIP25539, HIP26640 \cite{HIP}). We studied these cases with our method, using the corrections calculated with the theoretical point spread function (PSF) adopted for the ray-tracing, as the real PSF was proved to be consistent with it via direct measurements \cite{astri_optical_validation}. In principle, each star path should lie along a circular trajectory, but we fitted the data with arches of ellipse, in order to account for possible distortions due for example to a drift in the tracking. Moreover, we found that it is fundamental to fit simultaneously all the four stars considered, with a unique multi-ellipse function that we defined, otherwise the output will be contaminated by the residual effects of large pixelization. The parameters of our 4-stars fit were the following:
\begin{itemize}
    \item center coordinates $(X0, Y0)$;
    \item rotation angle $\theta $ of the ellipse;
    \item eccentricity;
    \item semi-major axes of stars.
\end{itemize}
Among them, we were particularly interested in the coordinates of the center, as it expresses the position of the telescope optical axis. We found out that in none of the 3 runs considered the center was compatible with the geometric center of the camera and hence we considered the average and the standard deviation of the 3 results as the best estimator for the real position of the telescope optical axis in the camera and its error. The final situation is illustrated in the green box of figure~\ref{fig_5}. A quantitative description of these results will be provided soon in a specific publication in preparation~\cite{Iovenitti2021}.

\section{Conclusion}
\noindent
The procedure presented in this contribution constitutes a powerful strategy to enhance the pointing accuracy of the ASTRI telescope, without any additional hardware.
In fact, it is possible to asses with it the alignment of the Cherenkov camera with the optical axis of the telescope, a quantity very difficult to be measured otherwise. For this reason, in the context of the incoming ASTRI Mini-Array this procedure will be exploited during both the assembly integration verification and the calibration phase \cite{Mineo2021}. Actually, it constitutes a method that can be easily replicated on every long observing run, constantly monitoring the accuracy of the pointing.

\section*{Acknowledgements}
\noindent
This work was conducted in the context of the ASTRI Project. This work is supported by the Italian Ministry of University and Research (MUR) with funds specifically assigned to the Italian National Institute for Astrophysics (INAF). We acknowledge support from the Brazilian Funding Agency FAPESP (Grant 2013/10559-5) and from the South African Department of Science and Technology through Funding Agreement 0227/2014 for the South African Gamma-Ray Astronomy Programme. This work has been supported by H2020-ASTERICS, a project funded by the European Commission Framework Programme Horizon 2020 Research and Innovation action under grant agreement n. 653477. IAC is supported by the Spanish Ministry of Science and Innovation (MICIU).

\end{document}